\documentclass[12pt,tightenlines,eqsecnum,floats,aps,amsmath,amssymb,nofootinbib,prd,showpacs,superscriptaddress]{revtex4}

\usepackage{amsmath,amssymb,amsfonts,amsthm} 

\usepackage{hyperref} 

\def\be{\begin{equation}}
\def\ee{\end{equation}}
\def\ba{\begin{eqnarray}}
\def\ea{\end{eqnarray}}
\def\nn{\nonumber}

\def\a{\alpha}

\def\g{\gamma}
\def\d{\delta}
\def\e{\epsilon}

\def\z{\zeta}

\def\l{\lambda}
\def\m{\mu}

\def\x{\xi}
\def\p{\pi}

\def\r{\rho}

\def\ph{\phi}

\def\ps{\psi}
\def\o{\omega}

\def\D{\Delta}

\def\Ps{\Psi}
\def\O{\Omega}

\newcommand{\cyl}{{\rm Cyl}} 
\newcommand{\hilbert}{{\cal H}} 
\newcommand{\sw}{{\cal S}} 
\newcommand{\dual}[1]{{#1}^\star} 
\newcommand{\norm}[1]{{\Vert{#1}\Vert}} 
\newcommand{\inner}[2]{{\langle {#1}\vert {#2} \rangle}} 
\newcommand{\gsp}[2]{{( {#1}\vert {#2} \rangle}} 
\newcommand{\ket}[1]{\vert{#1}\rangle} 
\newcommand{\bra}[1]{\langle{#1}\vert} 
\newcommand{\round}[1]{({#1}\vert} 

\newcommand{\opU}[1]{\hat{U}({#1})} 
\newcommand{\opV}[1]{\hat{V}({#1})} 

\newcommand{\secref}[1]{Section~\ref{#1}}

\newcommand{\eqnref}[1]{(\ref{#1})}

\begin{document}


\title{Galileo Symmetries in Polymer Particle Representation}
\author{Dah-Wei Chiou}
\affiliation{Department of
Physics, University of California, Berkeley, CA 94720,
USA}
\affiliation{Institute for Gravitational Physics and Geometry, Department
of Physics, The Pennsylvania State University, University Park, PA
16802, USA}

\begin{abstract}
To illustrate the conceptual problems for the low-energy symmetries
in the continuum of spacetime emerging from the discrete quantum
geometry, Galileo symmetries are investigated in the polymer
particle representation of a non-relativistic particle as a simple
toy model. The complete Galileo transformations (translation,
rotation and Galileo boost) are naturally defined in the polymer
particle Hilbert space and Galileo symmetries are recovered with
highly suppressed deviations in the low-energy regime from the
underlying polymer particle description.
\end{abstract}


\maketitle


\section{Introduction}
The non-perturbative, background-independent approach to the quantum
theory of general relativity has been successfully formulated in the
loop quantum gravity (LQG) and many of the long-standing challenges
are resolved. However, the non-perturbative feature of
background independence makes it very difficult to relate the spin
network states of LQG to the Fock states normally used in the
perturbative field theory for the low-energy physics, leaving the
important question unsolved: What is the connection between the
discrete geometry of LQG and the smooth structure of physical
geometry we perceive around us? To bridge the gap between the
background-independent quantum gravity and the familiar low-energy
physics, the analysis of semiclassical issues was proposed in
\cite{Varadarajan:1999aj}. The relation between polymer and Fock
excitations was then investigated in
\cite{Varadarajan:1999it,Varadarajan:2001nm,Ashtekar:2001xp} for
Maxwell field and in \cite{Varadarajan:2002ht} for linearized
gravity.

Based on the key ideas introduced in \cite{Ashtekar:2001xp}, to
illustrate the conceptual problems through a toy model, the
\emph{polymer particle representation} of quantum mechanics of a
non-relativistic particle was constructed in \cite{Ashtekar:2002sn},
where it is demonstrated that the relevant low-energy physics can be
extracted from elements of $\dual{\cyl}$ (the dual of the space of
the cylindrical functions of connections) by examining their
\emph{shadows} in the polymer particle Hilbert space $\hilbert_{\rm
Poly}$, even though $\dual{\cyl}$ does \emph{not} admit a natural
inner product.

Another related question regarding the low-energy description of LQG
is: How and in what precise sense does the low-energy symmetry in
the continuum of spacetime (e.g. Poincar\'{e} symmetry in Minkowski
space) arise from the discrete geometry of the full theory? (For the
attempt to derive the \emph{Minkowski vacuum} in
background-independent quantum gravity, see \cite{Conrady:2003en}.
See also \cite{Velhinho:2001qj} for the invariance of $U(1)$
holonomies.) To illuminate the conceptual issues about low-energy
symmetries, the polymer particle representation of quantum mechanics
constructed in \cite{Ashtekar:2002sn} can again serve as a simple
toy model. In this paper, we study the non-relativistic quantum
mechanics and thus focus on the Galileo symmetries. The notions of
Galileo transformations (translation, rotation and Galileo boost)
are naturally defined in the polymer particle Hilbert space via the
\emph{polymer coherent states}. On top of these notions, the Galileo symmetries and
their consequences (e.g. conservation laws) are investigated in the
polymer particle description, showing that the Galileo symmetries in
the Schr\"{o}dinger representation of the standard quantum mechanics
are recovered with negligible deviations in the low-energy regime.
The investigation also suggests that the temporal discreteness may
somehow descend from the spatial discreteness when the dynamics is
probed by a boosted reference.

This paper is organized as follows. In \secref{sec:outline}, the formulation of the polymer particle description is briefly outlined. \secref{sec:transformations} is devoted to the definition and analysis of the finite Galileo transformations in the polymer particle framework. \secref{sec:classical} then shows that, when probed on shadow states, these transformations reproduce the desired classical results in the appropriate low-energy approximation. The dynamical aspects of the Galileo symmetries in the standard and polymer quantum mechanics are investigated in \secref{sec:symmetries}. The results and their implications are summarized in \secref{sec:conclusion}.

\section{Outline of the Polymer Particle Description}\label{sec:outline}
The formulation and notations of the polymer particle framework are
outlined in this section; the details can be found in
\cite{Ashtekar:2002sn}. The real line $\mathbb{R}$ on which the
particle moves is now straightforwardly extended to $\mathbb{R}^3$
to include the rotation.

In the polymer particle representation, $\cyl_\g$ is the space of
complex-valued functions $f(\vec{k})$ of the type
\be
f(\vec{k})=\sum_j f_j\,e^{-i\vec{x}_j\cdot\vec{k}}
\ee
on $\mathbb{R}^3$, where $f_j$ are complex numbers with a suitable
fall-off and the \emph{graph} $\g$ consists of a countable set $\{\vec{x}_j\}$ of points on $\mathbb{R}^3$
with some technical conditions.\footnote{Aside from countability, the set $\{\vec{x}_j\}$ has to satisfy two additional conditions,
which ensure convergence of certain series of $\{\vec{x}_j\}$. See \cite{Ashtekar:2002sn} for the details.}
The functions $f(\vec{k})$ is said to be \emph{cylindrical} with respect
to $\g$.

Let $\cyl:=\bigcup_\g\cyl_\g$ for \emph{all} possible graphs; $\cyl$
is the infinite-dimensional vector space of functions on
$\mathbb{R}^3$ which are cylindrical with respect to some graph. The
\emph{polymer particle Hilbert space} $\hilbert_{\rm Poly}$
underlying the polymer particle representation is the Cauchy
completion of $\cyl$ with respect to the Hermitian inner product on
$\cyl$: \be
\inner{e^{-i\vec{x}_i\cdot\vec{k}}}{e^{-i\vec{x}_j\cdot\vec{k}}}
=\d_{\vec{x}_i,\vec{x}_j}, \ee where the right-hand side is the
Kronecker $\d$ (not Dirac distribution) and the uncountable basis
$\{\ket{e^{-i\vec{x}_i\cdot\vec{k}}}\}$ of $\cyl$ is labeled by
arbitrary real vectors $\vec{x}_i$.

The Weyl-Heisenberg algebra is represented on $\hilbert_{\rm Poly}$
as:
\be
\hat{W}(\vec{\z})f(\vec{k})=
\left[e^{\frac{i}{2}\vec{\l}\cdot\vec{\m}}\,\opU{\vec{\l}}\opV{\vec{\m}}\right]
f(\vec{k})
\qquad
\text{for}
\quad
\vec{\z}:=d_0\vec{\l}+i(\vec{\m}/d_0),
\ee
where the length scale $d_0$ is introduced to make $\vec{\z}$ dimensionless and the actions of $\hat{U}$ and $\hat{V}$
are given by
\be
\opU{\vec{\l}}f(\vec{k})=f(\vec{k}-\vec{\l})
\quad \text{and} \quad
\opV{\vec{\m}}f(\vec{k})=e^{i\vec{\m}\cdot\vec{k}}f(\vec{k}).
\ee
When acting on the orthonormal basis $\{\ket{\vec{x}_i}\}$,
the fundamental operators $\hat{U}$ and $\hat{V}$ yield
\be\label{eqn:U and V}
\opU{\vec{\l}}\ket{\vec{x}_i}=e^{i\vec{\l}\cdot\vec{x}_i}\ket{\vec{x}_i}
\quad \text{and} \quad
\opV{\vec{\m}}\ket{\vec{x}_i}=\ket{\vec{x}_i-\vec{\m}}.
\ee
The position operator $\hat{{\vec x}}$ is also well-defined:
\be
\hat{{\vec x}}\ket{{\vec x}_i}={\vec x}_i\ket{{\vec x}_i}.
\ee
In contrast, the momentum operator $\hat{\vec p}\equiv\hbar\hat{\vec k}$ does \emph{not}
exist; instead, $\opV{\vec \m}$ is used as the ``holonomy'' of the momentum.
($\hat{\vec{x}}$ is analogous to the electric flux operators and its eigenstates
provides the analogue of spin network states; $\opV{\vec{\m}}$ is analogous to
the holonomy of connections.)

The (algebraic) dual of $\cyl$ is denoted as $\dual{\cyl}$ and the
inclusions give the ``Gel'fand-type'' triplet:
\be
\cyl\subset\hilbert_{\rm Poly}\subset\dual{\cyl}.
\ee
The Schwartz space $\sw$ of the Schr\"{o}dinger Hilbert space $\hilbert_{\rm Sch}$ is
embedded in $\dual{\cyl}$. The fact that $\sw\subset\dual{\cyl}$ is used to define
a unique element $\round{\Ps}$ in $\dual{\cyl}$ for
each element $\ket{\ps}\in\sw$.
In particular, consider the standard coherent state $\ket{\ps_{\vec{\z_0}}}$
given by
\be\label{eqn:coherent state}
\ps_{\vec{\z}_0}(\vec{x})\equiv\psi_{\vec{x}_0,\vec{p}_0}(\vec{x})
=c\,e^{-\frac{(\vec{x}-\vec{x}_0)^2}{2d^2}}
e^{i\vec{k}_0\cdot(\vec{x}-\vec{x}_0)}
\qquad\text{for}\quad
\vec{\z}_0:=\frac{1}{\sqrt{2}\,d}(\vec{x}_0+i\vec{k}_0d^2),
\ee
which is peaked at
$\vec{x}_0$ for position $\hat{\vec{x}}$ and $\vec{p}_0\equiv\hbar\vec{k}_0$ for momentum $\hat{\vec{p}}$. To specify the ``tolerance'' (i.e. the uncertainties in $x_i$ and $p_i$) of the coherent state, the length scale $d$ is introduced to give the uncertainty $\D{x}_i=d/\sqrt{2}$ in $\hat{x}_i$ and $\D{p}_i=\hbar/(\sqrt{2}d)$ in $\hat{p}_i$.\footnote{In the case of a harmonic oscillator, $d$ is generally taken to be $\sqrt{\hbar/(m\o)}$.}${}^,$\footnote{In fact, we could consider the more generic form of coherent states $\ket{\ps_{\vec{x}_0,\vec{p}_0,{\bf A}}}$
given as $\ps_{\vec{x}_0,\vec{p}_0,{\bf A}}(\vec{x})=
c\,e^{-\frac{1}{2}(\vec{x}-\vec{x}_{0})^T{\bf A}^{-1}(\vec{x}-\vec{x}_{0})}
e^{i\vec{k}_0\cdot(\vec{x}-\vec{x}_0)}$ with an arbitrary
(positive-definite and symmetric) matrix ${\bf A}$ for the covariance of $\hat{x}_i$
(and $\hbar^2{\bf A}^{-1}$ for the covariance of $\hat{p}_i$).
All the results discussed in this paper can be easily extended
for the generic case and thus for simplicity we only consider
the restricted states given by \eqnref{eqn:coherent state},
in which ${\bf A}=d^2{\bf 1}_{3\times 3}$.\label{note:coherent state}}
The standard coherent state $\ket{\ps_{\vec{\z_0}}}$ uniquely
gives the corresponding \emph{polymer coherent state}:
\be
\round{\Ps_{\vec{\z_0}}}
=\bar{c}\sum_{\vec x}\left[e^{-\frac{(\vec{x}-\vec{x}_0)^2}{2d^2}}
e^{-i\vec{k}_0\cdot(\vec{x}-\vec{x}_0)}\right]\round{\vec x},
\ee
where $c$ is a normalization constant and $\{\round{\vec{x}}\}$
is a basis of $\dual{\cyl}$ dual to $\{\ket{\vec{x}_i}\}$ defined
in the obvious fashion:
\be
\gsp{\vec{x}}{\vec{x}_j}=\d_{\vec{x},\vec{x}_j}.
\ee
The juxtaposed notation $\gsp{\Ps}{f}$ denotes
the action of $\round{\Ps}\in\dual{\cyl}$ on $\ket{f}\in\cyl$;
i.e., $\round{\Ps}$ maps $\ket{f}$ to the complex number $\gsp{\Ps}{f}$.

In the polymer particle framework, the momentum operators
$\hat{p}_i$ do \emph{not} exist. Therefore, to approximate
$\hat{p}_i$ of the Schr\"{o}dinger representation, the ``fundamental
operators'' $\hat{K}_{i,\m_0}$ on $\hilbert_{\rm Poly}$ are defined
by introducing the \emph{fundamental length} $\m_0$.
Correspondingly, the angular momentum operators $\hat{L}_i$ are
replaced by $\hat{L}_{i,\m_0}$. These are discussed in
\secref{sec:commutation}.

Also note that $\dual{\cyl}$ does \emph{not} admit a natural inner
product, which make it difficult to define expectation values for
physical observables. The strategy to extract physical information
from the elements $\round{\Ps}$ of $\dual{\cyl}$ by the use of
\emph{shadow states} will be addressed in \secref{sec:classical}.

\section{The Complete Galileo Transformations}\label{sec:transformations}
In this section, we study the complete Galileo transformations in the
polymer particle framework including the spatial translation,
rotation and the pure Galileo transformation (Galileo boost).

\subsection{Defining Transformations in the Polymer Particle Framework}\label{sec:defining transform}
Continuous transformations are well studied in the standard
Schr\"odinger framework. In particular, spatial translation,
rotation and Galileo transformation can be readily realized in the
Schr\"odinger Hilbert space $\hilbert_{\rm Sch}$ by requiring the
expectation values of the position and momentum behave the same as
the transformations act on the classical phase space. To study the
transformations in polymer particle Hilbert space $\hilbert_{\rm
Poly}$, however, we cannot follow the standard strategy, since the
momentum operator $\hat{\vec{p}}$ fails to be well-defined in
$\hilbert_{\rm Poly}$. Furthermore, the transformation cannot be
defined by specifying its infinitesimal action because the
associated generator does not exist.

Nevertheless, we can exploit the fact that $\sw$ is embedded in
$\dual{\cyl}$ to extract the information of transformations in
the polymer particle framework. In \cite{Ashtekar:2002sn}, it is
shown that each element $\ket{\ps}\in\sw$ defines a unique element
$\round{\Ps}$ in $\dual{\cyl}$; in particular, the standard coherent
state $\ket{\ps_{\vec{\z_0}}}$ gives the polymer coherent state
$\round{\Ps_{\vec{\z_0}}}$.

If the finite transformation $\hat{\O}$ maps
$\ket{\ps_{\vec{\z_0}}}$ to $\ket{\ps_{\vec{\z'_0}}}$, which is
again a coherent state,\footnote{We will see later that it is the
case for the complete Galileo transformations.} then we can define the
operation of $\hat{\O}$ on $\dual{\cyl}$ via\footnote{The daggered
notation here merely means $\hat{\O}$ acts on the left and satisfies
$\round{a\Ps_1+b\Ps_2}\hat{\O}^\dagger=
\bar{a}\,\round{\Ps_1}\hat{\O}^\dagger+\bar{b}\,\round{\Ps_2}\hat{\O}^\dagger$.
The notation is reminiscent of the
resemblance between the inner product $\inner{\,\cdot\,}{\,\cdot\,}$
and the action $\gsp{\,\cdot\,}{\,\cdot\,}$, but $\hat{\O}^\dagger$ is not the adjoint of $\hat{\O}$ (since there is
no inner product structure in $\dual{\cyl}$).}
\be\label{eqn:def1}
\hat{\O}:\round{\Ps_{\vec{\z_0}}} \longmapsto
\round{\Ps'_{\vec{\z_0}}}\equiv
\round{\Ps_{\vec{\z_0}}}\hat{\O}^\dagger=\round{\Ps_{\vec{\z'_0}}}.
\ee
Equipped with the transformation on the polymer coherent state:
$\round{\Ps}\rightarrow\round{\Ps'}=\round{\Ps}\hat{\O}^\dagger$, we
can then define the operation of $\hat{\O}$ on any state of $\cyl$:
$\ket{f}\rightarrow\ket{f'}=\hat{\O}\ket{f}$ by requiring
\be\label{eqn:def2}
\gsp{\Ps'}{f}=\gsp{\Ps}{f'} \quad \text{or}
\quad \left[\round{\Ps}\hat{\O}^\dagger\right]\ket{f}=
\round{\Ps}\left[\hat{\O}\ket{f}\right].
\ee
This strategy serves as
a sound definition to carry out the transformations in the polymer
particle description.

On the other hand, $\cyl$ itself carries one-parameter groups
$\opU{\vec{\l}}$ and $\opV{\vec{\m}}$ defined in \eqnref{eqn:U and
V}. We expect that some of the (finite) Galileo transformations in
$\cyl$ can be expressed in terms of $\opU{\vec{\l}}$ and
$\opV{\vec{\m}}$. In particular, $\opV{\vec{\m}}$ is in fact the
spatial translation operator. (Although this seems obvious, we will
prove it as a consequence of the above prescription.) In the
following subsections, we will derive the explicit expressions for translation,
rotation and the Galileo boost.

\subsection{Translation}\label{sec:translation}
In a classical system, the spatial translation is defined as
\be\label{eqn:translation classical}
\vec{x}\rightarrow\vec{x'}=\vec{x}+\vec{\m},\qquad
\vec{p}\rightarrow\vec{p'}=\vec{p}.
\ee
In the standard quantum mechanics, the corresponding translation operator gives
\be
\hat{T}(\vec{\m})\ket{\vec{x}}=\ket{\vec{x}+\vec{\m}}
\ee
or equivalently
\be
\hat{T}(\vec{\m})\ps(\vec{x})=\ps(\vec{x}-\vec{\mu}).
\ee

Given a coherent state of $\sw$ peaked at $\vec{x}_0$ and $\vec{p}_0$
as defined in \eqnref{eqn:coherent state},
under the translation, it is transformed via
\ba
\psi_{\vec{x}_0,\vec{p}_0}(\vec{x})&\rightarrow&
\psi'_{\vec{x}_0,\vec{p}_0}(\vec{x})=\hat{T}(\vec{\m})\psi_{\vec{x}_0,\vec{p}_0}(\vec{x})\nn\\
&=&
c\,e^{-\frac{(\vec{x}-\vec{\m}-\vec{x}_0)^2}{2d^2}}
e^{i\vec{p}_0\cdot(\vec{x}-\vec{\m}-\vec{x}_0)/\hbar}
=\psi_{\vec{x}_0+\vec{\m},\vec{p}_0}(\vec{x}),
\ea
which is again a coherent state peaked at $\vec{x}_0+\vec{\m}$ and $\vec{p}_0$ as expected.
Therefore, by \eqnref{eqn:def1}, translation acting on the polymer coherent state gives
\ba\label{eqn:translation in polymer}
\round{\Ps_{\vec{x}_0,\vec{p}_0}} \rightarrow \round{\Ps'_{\vec{x}_0,\vec{p}_0}}&=&
\round{\Ps_{\vec{x}_0,\vec{p}_0}}\hat{T}^\dagger(\vec{\m})=
\round{\Ps_{\vec{x}_0+\vec{\m},\vec{p}_0}}
\ea
or, in terms of the dual basis in $\dual{\cyl}$,
\be
\round{\vec{x}}\hat{T}^\dagger(\vec{\m})=\round{\vec{x}+\vec{\m}},
\ee
which together with \eqnref{eqn:def2} yields the translation acting on the basis of $\cyl$:
\be\label{eqn:translation}
\hat{T}(\vec{\m})\ket{\vec{x}_i}=\ket{\vec{x}_i+\vec{\m}}
\ee
by the identity $\gsp{\vec{x}}{\vec{x}_i}=\d_{\vec{x},\vec{x}_i}$.
As expected, this shows
\be
\hat{T}(\vec{\m})=\opV{-\vec{\m}}.
\ee

\subsection{Rotation}\label{sec:rotation}
In the classical phase space, rotation is specified by a rotational matrix ${\bf R}$ such that
\be\label{eqn:rotation classical}
\vec{x}\rightarrow \vec{x'}={\bf R}\vec{x}, \qquad \vec{p}\rightarrow \vec{p'}={\bf R}\vec{p}.
\ee
Under the rotation,
the standard coherent state is transformed via\footnote{As mentioned in Footnote~\ref{note:coherent state},
if we consider the generic coherent state $\ps_{\vec{x}_0,\vec{p_0},{\bf A}}(\vec{x})$,
we will have $\ps_{\vec{x}_0,\vec{p_0},{\bf A}}(\vec{x})\rightarrow\ps'_{\vec{x}_0,\vec{p_0},{\bf A}}(\vec{x})
=\ps_{{\bf R}\vec{x}_0,{\bf R}\vec{p_0},{\bf R}{\bf A}{\bf R}^{\!T}}(\vec{x})$ under rotation.
This still leads to the same result in \eqnref{eqn:rotation}.}
\ba
\psi_{\vec{x}_0,\vec{p}_0}(\vec{x})&\rightarrow&
\psi'_{\vec{x}_0,\vec{p}_0}(\vec{x})=\hat{D}({\bf R})\psi_{\vec{x}_0,\vec{p}_0}(\vec{x})\nn\\
&=&
c\,e^{-\frac{({\bf R}^{-1}\vec{x}-\vec{x}_0)^2}{2d^2}}
e^{i\vec{p}_0\cdot({\bf R}^{-1}\vec{x}-\vec{x}_0)/\hbar}
=\psi_{{\bf R}\vec{x}_0,{\bf R}\vec{p}_0}(\vec{x}),
\ea
from which we deduce the rotation acting on
$\dual{\cyl}$ and $\cyl$ as
\be
\round{\vec{x}}\rightarrow \round{\vec{x}}\hat{D}^\dagger({\bf R})=\round{{\bf R}\vec{x}}
\ee
and
\be\label{eqn:rotation}
\ket{\vec{x}_i}\rightarrow \hat{D}({\bf R})\ket{\vec{x}_i}=\ket{{\bf R}\vec{x}_i}
\ee
by following the same argument in \secref{sec:translation}.

Note that $\hat{D}({\bf R})$ cannot be expressed in terms of the
fundamental operators $\opU{\vec{\l}}$ and $\opV{\vec{\m}}$, unlike
the cases for translation and Galileo boost (the latter will be
studied in \secref{sec:galileo}). Nevertheless, the space $\cyl$
serves as a good carrier space and is closed for all finite group
elements of rotation, although the infinitesimal rotation and thus
the generator (angular momentum) cannot be represented in the
polymer particle framework, which will cause some problems when we
study the rotational symmetry in \secref{sec:rotational symmetry}.

\subsection{Galileo Boost}\label{sec:galileo}
The pure Galileo transformation (Galileo boost) of the standard
quantum mechanics is discussed in \cite{Gottfried}. We follow the
steps thereof and extend the notions to the polymer particle
framework.

Let $F$ be an inertial frame, in which the position and momentum of a particle are $\vec{x}$ and $\vec{p}$.
Given a second inertial frame $F'$, which at $t=0$ coincides with $F$ and is moving with velocity
$-\vec{v}$ as seen from $F$. The same particle will have the position $\vec{x}+\vec{v}\,t$ and momentum
$\vec{p}+m\vec{v}$ as seen from $F'$. Both frames have the same temporal lapse.
The Galileo transformation is defined in classical phase space as the boost:
\be\label{eqn:galileo classical}
\vec{x}\rightarrow\vec{x'}=\vec{x}+\vec{v}\,t, \qquad
\vec{p}\rightarrow\vec{p'}=\vec{p}+m\vec{v}, \qquad
t'=t,
\ee
where $m$ is the mass of the particle.

Because pure Galileo transformations form a continuous group, in
standard quantum mechanics, they are carried out by unitary
operators $\hat{G}(\vec{v},t)$ which are to produce\footnote{Here,
we treat $t$ as a pure parameter and thus consider a family of
Galileo transformations parameterized by $\vec{v}$ and $t$. We
freeze the state in the Hilbert space by dismissing $t$ as the
temporal variable in order to study the instantaneous transformation
at time $t$ and disregard the dynamics, which will be studied
in \secref{sec:galileo symmetry}.}
\be\label{eqn:galileo quantum}
\hat{G}^\dagger(\vec{v},t)\ \hat{\vec{x}}\ \hat{G}(\vec{v},t)=\hat{\vec{x}}+\vec{v}\,t,
\qquad
\hat{G}^\dagger(\vec{v},t)\ \hat{\vec{p}}\ \hat{G}(\vec{v},t)=\hat{\vec{p}}+m\vec{v}.
\ee
As with other continuous transformations, $\hat{G}$ can be written as the exponential of its generator
\be
\hat{G}(\vec{v},t)=e^{-i\vec{v}\cdot\hat{\vec{N}}(t)/\hbar},
\ee
where the Hermitian operator $\hat{\vec{N}}$ will be called the \emph{boost}.
By requiring the infinitesimal transformation $(1-i\d\vec{v}\cdot\!\hat{\vec{N}}/\hbar)$ to produce
\eqnref{eqn:galileo quantum}, we have
\be
\frac{i}{\hbar}[\hat{N}_i,\hat{x}_j]=\d_{ij}t, \quad
\frac{i}{\hbar}[\hat{N}_i,\hat{p}_j]=\d_{ij}m,
\ee
which gives
\be\label{eqn:boost generator}
\hat{\vec{N}}(t)=\hat{\vec{p}}\,t-m\hat{\vec{x}}.
\ee

By the Baker-Campbell-Hausdorff formula for the case of two operators $A$ and $B$ whose commutator is a
c-number: $e^A e^B=e^{A+B}e^{\frac{1}{2}[A,B]}$, \eqnref{eqn:galileo quantum} and
\eqnref{eqn:boost generator} lead to
\be\label{eqn:galileo transform}
\hat{G}(\vec{v},t)=e^{\frac{i}{\hbar}\frac{mv^2}{2}t}
e^{-t\vec{v}\cdot\hat{\vec{p}}/\hbar}e^{im\vec{v}\cdot\hat{\vec{x}}/\hbar}
\ee
or equivalently when acting on the wave functions
\be\label{eqn:galileo transform2}
\hat{G}(\vec{v},t):
\psi(\vec{x})\longmapsto\psi'(\vec{x})=
e^{im(\vec{v}\cdot\vec{x}-\frac{1}{2}v^2t)/\hbar}\psi(\vec{x}-\vec{v}t).
\ee
The result of \eqnref{eqn:galileo transform} and \eqnref{eqn:galileo transform2} suggests that
$\hat{G}(\vec{v},t)=e^{\frac{i}{\hbar}\frac{mv^2}{2}t}\ \opV{-\vec{v}t}\,\opU{m\vec{v}/\hbar}$
when it acts on $\cyl$, but again we will see that this is a consequence of the prescription
for defining transformations mentioned in \secref{sec:defining transform}.

Given a coherent state in \eqnref{eqn:coherent state},
the wave function is transformed under Galileo transformation by \eqnref{eqn:galileo transform2}:
\ba\label{eqn:galileo on coherent}
\psi_{\vec{x}_0,\vec{p}_0}(\vec{x})&\rightarrow&
\psi'_{\vec{x}_0,\vec{p}_0}(\vec{x})=\hat{G}(\vec{v},t)\psi_{\vec{x}_0,\vec{p}_0}(\vec{x})
=c\,e^{im(\vec{v}\cdot\vec{x}-\frac{1}{2}v^2t)/\hbar}
e^{-\frac{(\vec{x}-\vec{x}_0-\vec{v}t)^2}{2d^2}}
e^{i\vec{p}_0\cdot(\vec{x}-\vec{x}_0-\vec{v}t)/\hbar}\nn\\
&=&
c\,e^{im\vec{v}\cdot(\vec{x}_0+\frac{1}{2}\vec{v}t)/\hbar}
e^{-\frac{(\vec{x}-\vec{x}_0-\vec{v}t)^2}{2d^2}}
e^{i(\vec{p}_0+m\vec{v})\cdot(\vec{x}-\vec{x}_0-\vec{v}t)/\hbar}\nn\\
&=&e^{im\vec{v}\cdot(\vec{x}_0+\frac{1}{2}\vec{v}t)/\hbar}
\psi_{\vec{x}_0+\vec{v}t,\vec{p}_0+m\vec{v}}(\vec{x}),
\ea
which is again a coherent state.\footnote{
The reason why coherent states remain coherent under
Galileo transformation can be understood as follows. Compute the commutator of the boost and annihilation
operators: $[\hat{a}_i,\hat{N}_j]=\frac{\d_{ij}}{\sqrt{2}d}(i\hbar t-d^2m$), where the annihilation operator
$\hat{\vec{a}}=\frac{1}{\sqrt{2}d}(\hat{\vec{x}}+i\frac{d^2}{\hbar}\hat{\vec{p}})$. The identity
$[\hat{B},e^{x\hat{A}}]=e^{x\hat{A}}[\hat{B},\hat{A}]x$
then leads to
$[\hat{a}_i, \hat{G}(\vec{v},t)]=\frac{v_i}{\sqrt{2}d}(t+\frac{i}{\hbar}d^2m)\hat{G}(\vec{v},t)$.
If $\hat{\vec{a}}\ket{\ps_{\vec{\z}_0}}=\vec{\z}_0\ket{\ps_{\vec{\z}_0}}$, consequently
$\hat{\vec{a}}\,\hat{G}(\vec{v},t)\ket{\ps_{\vec{\z}_0}}
=\left(\vec{\z}_0+\frac{\vec{v}}{\sqrt{2}d}(t+\frac{i}{\hbar}d^2m)\right)
\hat{G}(\vec{v},t)\ket{\ps_{\vec{\z}_0}}$.
Therefore, $\hat{G}(\vec{v},t)\ket{\ps_{\vec{\z}_0}}$ is again an eigenstate of $\hat{\vec{a}}$ with
the eigenvalue $\vec{\z}'_0=\vec{\z}_0+\frac{\vec{v}}{\sqrt{2}d}(t+\frac{i}{\hbar}d^2m)
=\frac{1}{\sqrt{2}d}\left((\vec{x}_0+\vec{v}t)+i\frac{d^2}{\hbar}(\vec{p}+m\vec{v})\right)$ as expected.
Similar argument can be applied to translation and rotation as well.
}

By \eqnref{eqn:def1}, the result of \eqnref{eqn:galileo on coherent} gives the transformation on the
polymer coherent state as
\be
\round{\Ps_{\vec{x}_0,\vec{p}_0}} \rightarrow \round{\Ps'_{\vec{x}_0,\vec{p}_0}}=
\round{\Ps_{\vec{x}_0,\vec{p}_0}}\hat{G}^\dagger(\vec{v},t)=
e^{-im\vec{v}\cdot(\vec{x}_0+\frac{1}{2}\vec{v}t)/\hbar}
\round{\Ps_{\vec{x}_0+\vec{v}t,\vec{p}_0+m\vec{v}}}
\ee
or, in terms of the dual basis in $\dual{\cyl}$,
\be
\round{\vec{x}}\hat{G}^\dagger(\vec{v},t)=
e^{-im\vec{v}\cdot(\vec{x}+\frac{1}{2}\vec{v}t)/\hbar}
\round{\vec{x}+\vec{v}t}.
\ee
According to \eqnref{eqn:def2} and the identity $\gsp{\vec{x}}{\vec{x}_i}=\d_{\vec{x},\vec{x}_i}$,
we obtain the Galileo transformation on the basis of $\cyl$:
\be
\hat{G}(\vec{v},t)\ket{\vec{x}_i}=e^{im\vec{v}\cdot(\vec{x}_i+\frac{1}{2}\vec{v}t)/\hbar}
\ket{\vec{x}_i+\vec{v}t}.
\ee
Compared with \eqnref{eqn:U and V}, the transformation can be written in terms of $\hat{U}$ and $\hat{V}$
as expected:
\be\label{eqn:G}
\hat{G}(\vec{v},t)=e^{\frac{i}{\hbar}\frac{mv^2}{2}t}\ \opV{-\vec{v}t}\,\opU{m\vec{v}/\hbar}.
\ee

\subsection{Commutation Relations of the Modified Galileo Algebras}\label{sec:commutation}
We have shown that the \emph{finite} transformations of the
complete Galileo transformations are well-defined in the polymer
particle framework. However, the infinitesimal transformations are
ill-defined and the corresponding generators (momentum, angular
momentum and boost) do \emph{not} exist in $\hilbert_{\rm Poly}$. In
this subsection, we define the modified generators by introducing
the fundamental length $\m_0$ as suggested in \cite{Ashtekar:2002sn}
and study the commutation relations of the modified Galileo
algebras.

The commutator between the basic operators $\hat{x}_i$ and $\opV{\m
\vec{e}_j}$ is
\be\label{eqn:x and V}
[\hat{x}_i,\opV{\m \vec{e}_j}]=-\m\,\d_{ij}\opV{\m
\vec{e}_j}.
\ee
On the other hand, the momentum operator is  not well-defined in
polymer particle description. Instead, we introduce the \emph{fundamental length}
$\m_0$, which is to be sufficiently small ($\ell\ll\m_0\ll d$)\footnote{As defined
in \cite{Ashtekar:2002sn}, $\ell$ is the spacing of the regular lattices on which
the \emph{shadow states} are defined, and
$d$ is the characteristic length scale which defines our tolerance.}, and define the analogue
of the momentum operator in $\hilbert_{\rm Poly}$ as
$\hbar \hat{\vec{K}}_{\!\m_0}$ with
\be\label{eqn:momentum operator}
\hat{K}_{i,\m_0}:=\frac{1}{2\m_0
i}\left(\hat{V}(\m_0 \vec{e}_i)-\hat{V}(-\m_0 \vec{e}_i)\right),
\ee
where $\vec{e}_i$ are the unit vectors in $x$-, $y$- or $z$-direction.
The commutator between the position and momentum becomes
\be\label{eqn:x and K}
[\hat{x}_i,\hat{K}_{j,\m_0}]=
\frac{i}{2}\d_{ij}\left(\opV{\m_0\vec{e}_j}+\opV{-\m_0\vec{e}_j}\right)
\equiv i\d_{ij}\hat{\x}_{j,\m_0},
\ee
where we define
\be
\hat{\x}_{j,\m_0}:=\frac{\opV{\m_0\vec{e}_j}+\opV{-\m_0\vec{e}_j}}{2}
\ee
and note that
$\hat{\x}_{j,\m_0}\rightarrow 1$ when $\m_0\rightarrow 0$ as the space becomes continuous.
By \eqnref{eqn:x and V}, We have
\be\label{eqn:x and xi}
[\hat{x}_i,\hat{\x}_{j,\m_0}]=-i\m_0^2\d_{ij}\hat{K}_{j,\m_0}.
\ee
Meanwhile, we also list the obvious commutation relations:
\be\label{eqn:xx and KK}
[\hat{x}_i,\hat{x}_j]=0,
\quad
[\hat{K}_{i,\m_0},\hat{K}_{j,\m_0}]=0,
\quad
[\hat{K}_{i,\m_0},\hat{\x}_{j,\m_0}]=0
\quad\text{and}\quad
[\hat{\x}_{i,\m_0},\hat{\x}_{j,\m_0}]=0
\ee

In the same spirit, the modified angular momentum for the rotation is defined in the similar way:
\be\label{eqn:angular momentum operator}
\hat{L}_{i,\m_0}:=\hbar\sum_{j,k}\e_{ijk}\,\hat{x}_j\hat{K}_{k,\m_0}.
\ee
By \eqnref{eqn:x and K}, \eqnref{eqn:x and xi} and \eqnref{eqn:xx and KK},
we then have\footnote{\textit{Note}: The index $j$ is not summed (no Einstein convention).}
\be\label{eqn:L and x}
[\hat{L}_{i,\m_0},\hat{x}_j]=i\hbar\sum_{k}\e_{ijk}\hat{x}_k\,\hat{\x}_{j,\m_0},
\ee
\be\label{eqn:L and K}
[\hat{L}_{i,\m_0},\hat{K}_{j,\m_0}]=i\hbar\sum_{k}\e_{ijk}\hat{K}_{k,\m_0}\,\hat{\x}_{j,\m_0},
\ee
\be\label{eqn:L and L}
[\hat{L}_{i,\m_0},\hat{L}_{j,\m_0}]=i\hbar\sum_{k}\e_{ijk}
\hat{L}_{k,\m_0}\hat{\x}_{k,\m_0}
\ee
and
\be\label{eqn:L and xi}
[\hat{L}_{i,\m_0},\hat{\x}_{j,\m_0}]=-i\hbar\m_0^2\hat{K}_{j,\m_0}\sum_{k}\e_{ijk}\hat{K}_{k,\m_0}.
\ee

Finally, analogous to \eqnref{eqn:boost generator}, we defined the
modified boost generator as \be\label{eqn:boost operator}
\hat{N}_{i,\m_0}(t):=\hbar\hat{K}_{i,\m_0}t-m\hat{x}_i. \ee This
leads to \be\label{eqn:K and N}
[\hat{K}_{j,\m_0},\hat{N}_{j,\m_0}(t)]=im\,\d_{ij} \ee and
\be\label{eqn:L and N}
[\hat{L}_{i,\m_0},\hat{N}_{j,\m_0}(t)]=i\hbar\sum_{k}\e_{ijk}\hat{N}_{k,\m_0}(t)\,\hat{\x}_{j,\m_0}.
\ee

The algebra of the modified generators of the complete Galileo group
is the same as its counterpart in the standard quantum mechanics if
we take the fundamental length $\m_0$ to be vanishing (i.e. the continuous
limit).

Finally, we also define
\be\label{eqn:L2}
\widehat{\vec{L}^2}_{\!\m_0}:=\hat{L}_{x,\m_0}^2+\hat{L}_{y,\m_0}^2+\hat{L}_{z,\m_0}^2,
\ee
and we have
\be\label{eqn:L2 and L}
[\widehat{\vec{L}^2}_{\!\m_0},\hat{L}_{z,\m_0}]=i\hbar
\left(\hat{L}_{x,\m_0}\hat{L}_{y,\m_0}+\hat{L}_{y,\m_0}\hat{L}_{x,\m_0}\right)
\left(\hat{\x}_{x,\m_0}-\hat{\x}_{y,\m_0}\right)
\ee
and so on. Thus, $\widehat{\vec{L}^2}_{\!\m_0}$ does not commute with $\hat{L}_{i,\m_0}$,
unless in the continuous limit.

Note that the position operator $\hat{x}_i$, modified momentum
$\hat{K}_{i,\m_0}$, and modified boost $\hat{N}_{i,\m_0}(t)$ form a
closed algebra together with $\hat{\x}_{i,\m_0}$. In contrast,
the modified angular momenta $\hat{L}_{i,\m0}$ do not form a closed
algebra even with $\hat{x}_i$, $\hat{K}_{i,\m_0}$,
$\hat{N}_{i,\m_0}(t)$ and/or $\hat{\x}_{i,\m_0}$ included, since the
right-hand sides of \eqnref{eqn:L and x}--\eqnref{eqn:L and xi} and
\eqnref{eqn:L and N} are not the sum of generators but the sum of
the products of two operators.

To reflect the discreteness of area and volume in LQG, as a toy
model, we impose the fundamental length $\m_0$ in the polymer
particle framework. The way the fundamental length is imposed is
however very specific: We modify the momentum operator as in
\eqnref{eqn:momentum operator} by introducing $\m_0$ only in the
three orthogonal directions ($x$-, $y$- and $z$-directions). The
rotational transformation does not leave this particular set-up
invariant. This is the underlying reason why the modified angular
momenta do not give a closed algebra. Related to this issue, when the
symmetry is taken into account, we will see that the rotational
symmetry is respected only approximately while the translational
symmetry still holds exactly in the polymer particle description.
(This will be the topic in \secref{sec:translational symmetry} and
\secref{sec:rotational symmetry}.)

\section{Classical Correspondence on the Shadow States}\label{sec:classical}
We have successfully defined the complete Galileo transformation in
the polymer particle framework. A natural question is to ask whether
these transformations defined in the polymer particle description
reproduce the desired classical behaviors, namely, whether
\eqnref{eqn:translation classical}, \eqnref{eqn:rotation classical}
and \eqnref{eqn:galileo classical} will hold as some kind of
classical limit when we focus on the polymer coherent states.

To circumvent the difficulty that the operator $\hat{\vec{p}}$ is
not defined in the polymer particle framework and there is no inner
product in $\dual{\cyl}$, we follow the strategy introduced in
\cite{Ashtekar:2002sn}. To begin with, define the \emph{shadow
state} $\ket{\Ps_\g^{\rm shad}}$ as the element $\round{\Ps}$ of
$\dual{\cyl}$ projected to the subspace $\cyl_\g$ by the projection
operator $\hat{p}_\g$:
\be
\round{\Ps}\hat{p}_\g
:= \sum_{\vec{x}_i\in\g}\Ps(\vec{x}_i)\ket{\vec{x}_i}
\equiv\ket{\Ps_\g^{\rm shad}},
\ee
with $\round{\Ps}=\sum_{\vec{x}}{\bar \Ps}(\vec{x})\round{\vec{x}}$.
Then, we say that the ``expectation values'' of the operator $\hat{A}$ is given by
\be
\langle\hat{A}\rangle=\frac{\round{\Ps}\hat{A}\ket{\Ps_\g^{\rm shad}}}{\norm{\Ps_\g^{\rm shad}}^2},
\ee
when the state $\round{\Ps}$ is probed by a sufficiently refined graph $\g$. We restrict ourselves to the shadows
on the regular lattices with sufficiently small lattice spacing $\ell$ ($\ell\ll d$).
We also adopt \eqnref{eqn:momentum operator} as the momentum operator.

The expectation values $\langle\hat{\vec{x}}\rangle$ and $\langle\hat{\vec{p}}\rangle$ have been investigated
for polymer coherent states in \cite{Ashtekar:2002sn}. The results are
\be\label{eqn:x expectation}
\langle\hat{x}_i\rangle_{\vec \z}:=
\frac{\round{\Ps_{\vec \z}}\hat{x}_i\ket{\Ps_{{\vec \z},\g}^{\rm shad}}}{\norm{\Ps_{{\vec \z},\g}^{\rm shad}}^2}
=x_i\left[1+{\cal O}\left(e^{-\frac{\p^2d^2}{\ell^2}}\right)\right]
\ee
and
\be\label{eqn:p expectation}
\langle\hat{p}_i\rangle_{\vec \z} \equiv \hbar\langle\hat{K}_{i,\m_0}\rangle_{\vec \z}:=\hbar
\frac{\round{\Ps_{\vec \z}}\hat{K}_{i,\m_0}\ket{\Ps_{{\vec \z},\g}^{\rm shad}}}{\norm{\Ps_{{\vec \z},\g}^{\rm shad}}^2}
=p_i\left[1+{\cal O}\left(k^2\m_0^2\right)+{\cal O}\left(\frac{\ell^2}{d^2}\right)\right],
\ee
where $\round{\Ps_{\vec \z}}$ is peaked at ${\vec \z}=\frac{1}{\sqrt{2}d}({\vec x}+id^2{\vec k})$
and $p_i=\hbar k_i$.

We have shown that under the complete Galileo transformations the coherent states remain coherent and
peaked at the values as the classical counterparts transfer positions and momenta; i.e up to a phase factor,
$\round{\Ps_{{\vec x},{\vec p}}} \rightarrow \round{\Ps_{\vec{x'},\vec{p'}}}$ with
the corresponding transformation in the classical
phase space $({\vec x},{\vec p}) \rightarrow (\vec{x'},\vec{p'})$.
Therefore, the transformation of the expectation values is given by
$\langle\hat{x}_i\rangle_{{\vec x},{\vec p}} \rightarrow \langle\hat{x}_i\rangle_{\vec{x'},\vec{p'}}$
and
$\langle\hat{p}_i\rangle_{{\vec x},{\vec p}} \rightarrow \langle\hat{p}_i\rangle_{\vec{x'},\vec{p'}}$,
which
are \emph{different} from its classical counterpart $({\vec x},{\vec p}) \rightarrow (\vec{x'},\vec{p'})$
according to \eqnref{eqn:x expectation} and \eqnref{eqn:p expectation}.

Although the expectation values do not reproduce the classical
result exactly, the discrepancy is highly negligible as long as
$\ell\ll d$ and $k\m_0\ll 1$. For $k\m_0\gtrsim 1$, however, the
transformation in polymer particle framework no longer agree with
the classical counterpart, but the non-relativistic approximation
breaks down long before one reaches such high momenta. Also note
that however small $k$ is, by the boost with a large ${\vec v}$, one
can always get a large momentum (${\vec k}'={\vec k}+m{\vec
v}/\hbar$) and hence the agreement fails to hold. This is again due
to the fact that we should take relativistic effect into account
when the boost velocity is very fast and therefore Galileo boost
should be replaced by Lorentz transformation.

It is remarkable to know that all the finite Galileo transformations
can be exactly represented in the polymer particle framework whereas
they agree with the classical results only approximately when the
expectation values are probed by refined graphs.

\section{Galileo Symmetries and Their Consequences}\label{sec:symmetries}
In standard quantum mechanics, a dynamic system is said to be
invariant or symmetric under a continuous transformation
$\hat{\O}(\r)$ if its Hamiltonian $\hat{H}$ remains unchanged by the
transformation; i.e.
$\hat{\O}(\r)^\dagger\hat{H}\,\hat{\O}(\r)=\hat{H}$ for all possible
values of $\r$, which parameterizes the transformation. There are
three important consequences if $\hat{H}$ respects a continuous
symmetry:
\begin{enumerate}
\item\label{consequence 1} If $\ket{\Ps(t)}$ satisfies the Schr\"odinger equation, so does $\ket{\Ps'(t)}:=\hat{\O}\ket{\Ps(t)}$.
\item\label{consequence 2} If $\hat{g}$ is the generator corresponding to the infinitesimal transformation,
it commutes with $\hat{H}$ and thus we can find the eigenstates both for energy and $\hat{g}$
simultaneously.
\item\label{consequence 3} The expectation value of $\hat{g}$ is conserved; i.e. $\bra{\Ps(t)}\,\hat{g}\,\ket{\Ps(t)}$
is a constant of motion.
\end{enumerate}

To study the dynamics in the polymer particle Hilbert space, due to
nonexistence of the momentum operator, we first replace the standard
Hamiltonian $\hat{H}=\frac{1}{2m}\widehat{{\vec
p}^{\,2}}+V(\hat{{\vec x}})$\footnote{\textit{Note}: Do not confuse
the potential $V(\hat{{\vec x}})$ with the fundamental operator
$\opV{\vec{\m}}$.} by the modified one in terms of
fundamental operators \cite{Ashtekar:2002sn}:
\be\label{eqn:hamiltonian}
\hat{H}_{\m_0}=\frac{\hbar^2}{2m}
\left(\widehat{K^2_x}_{\!\m_0}+\widehat{K^2_y}_{\!\m_0}+\widehat{K^2_z}_{\!\m_0}\right)+
V(\hat{{\vec x}})
\ee
with
\be\label{eqn:K square}
\widehat{K^2_i}_{\!\m_0}=\frac{1}{\m^2_0}\left(2-\opV{\m_0\vec{e}_i}-\opV{-\m_0\vec{e}_i}\right)
\equiv\frac{2}{\m_0^2}(1-\hat{\x}_{i,\m_0}),
\ee
and cast the Schr\"odinger equation in the polymer particle representation as
\be\label{eqn:schrodinger}
i\hbar\frac{d}{dt}\ket{f(t)}=\hat{H}_{\m_0}\ket{f(t)}
\quad\text{for}\quad
\ket{f(t)}\in\cyl.
\ee
If $\hat{H}_{\m_0}$ is independent of $t$, then the solution is
\be
\ket{f(t)}=e^{-i\hat{H}_{\m_0}t/\hbar}\ket{f(0)}.
\ee

Since the motivation to impose the fundamental length $\m_0$ is to
give the underlying discreteness analogous to that in quantum
geometry, we will adopt the viewpoint that the dynamics modified
with $\m_0$ is fundamental (whereas the continuous limit is
effective) and hence only $\hat{\vec{x}}$ and those $\opV{{\vec
\m}}$ with ${\vec \m}=\m_0(N_1{\vec e}_1+N_2{\vec e}_2+N_3{\vec
e}_3)$ for $N_i$ integers are observationally relevant. (See
\cite{Ashtekar:2002sn} for more details.)

With the fundamental dynamics defined, we can now say that the
system respects the continuous symmetry if and only if
\be\label{eqn:symmetry
criterion}
\hat{\O}^\dagger(\r)\hat{H}_{\m_0}\hat{\O}(\r)=\hat{H}_{\m_0} \ee
for all possible values of $\r$. It can be easily shown that
\eqnref{eqn:symmetry criterion} is the sufficient and necessary
condition for \be\label{eqn:dynamics with symmetry}
\hat{\O}^\dagger(\r)\,e^{-i\hat{H}_{\m_0}t/\hbar}\,\hat{\O}(\r)=e^{-i\hat{H}_{\m_0}t/\hbar}.
\ee
The meaning of \eqnref{eqn:dynamics with symmetry} is that
Consequence~\ref{consequence 1} is automatically satisfied in the polymer particle
representation once we adopt \eqnref{eqn:symmetry criterion} as the
definition of invariance.

On the other hand, opposed to the standard quantum mechanics,
\eqnref{eqn:symmetry criterion} does not necessarily imply
\be\label{eqn:g and H} [\hat{g}_{\m_0},\hat{H}_{\m_0}]=0 \ee where
$\hat{g}_{\m_0}$ is the associated modified generator
($\hat{g}_{\m_0}=\hbar \hat{K}_{i,\m_0}$ for translation and
$\hat{g}_{\m_0}=\hat{L}_{i,\m_0}$ for rotation). Hence, Consequence~\ref{consequence 2}
may not hold for the modified generator in the polymer particle
description.

In order to examine the analogue of Consequence~\ref{consequence 3}, again, we define the expectation value
with respect to the state $\round{\Ps}$ of $\dual{\cyl}$ probed on the graph $\g$ as
\be
\langle\hat{g}_{\m_0}\rangle_\g(t):=\round{\Ps(t)}\hat{g}_{\m_0}\ket{\Ps^{\rm shad}_\g(t)},
\ee
where
\be
\round{\Ps(t)}=\round{\Ps(0)}e^{i\hat{H}_{\m_0}t/\hbar}
\quad\text{and}\quad
\ket{\Ps^{\rm shad}_\g(t)}\equiv\round{\Ps(t)}\hat{p}_\g.
\ee
Note that $\round{\Ps(t)}$ may not remain as a coherent polymer state even if $\round{\Ps(0)}$ is.
Also notice that
$\ket{\Ps^{\rm shad}_\g(t)}\neq e^{-i\hat{H}_{\m_0}t/\hbar}\ket{\Ps^{\rm shad}_\g(0)}$
since $\hat{H}_{\m_0}$ and $\hat{p}_\g$ do not commute in general.

As discussed in \cite{Ashtekar:2002sn}, let $\a^{{\vec x}_0}$ be the regular lattice consisting of
points ${\vec x}_0+\m_0(N_1{\vec e}_1+N_2{\vec e}_2+N_3{\vec e}_3)$ with
${\vec x}_0\in[0,\m_0)^3$. The full polymer particle Hilbert space can be decomposed as a direct sum of
separable Hilbert spaces $\hilbert^{{\vec x}_0}_{\rm Poly}$ based on the graph $\a^{{\vec x}_0}$:
\be
\hilbert_{\rm Poly}=\bigoplus_{{\vec x}_0\in[0,\m_0)^3}\hilbert^{{\vec x}_0}_{\rm Poly}.
\ee
Since the observable algebra is now generated by $\hat{{\vec x}}$ and
$\opV{\m_0(N_1{\vec e}_1+N_2{\vec e}_2+N_3{\vec e}_3)}$, observables cannot mix states belonging to
distinct $\hilbert^{{\vec x}_0}_{\rm Poly}$ and each of these Hilbert spaces is
\emph{superselected}. Therefore, restricted to a single $\hilbert^{{\vec x}_0}_{\rm Poly}$,
$\hat{H}_{\m_0}$ and $\hat{p}_\g$ commute with each other and we then have
\be
\ket{\Ps^{\rm shad}_{{\vec x}_0}(t)}
=
e^{-i\hat{H}_{\m_0}t/\hbar}
\ket{\Ps^{\rm shad}_{{\vec x}_0}(0)}
\ee
and consequently
\be\label{eqn:time change of g}
\langle\hat{g}_{\m_0}\rangle_{{\vec x}_0}(t)=
\round{\Ps(0)}
e^{i\hat{H}_{\m_0}t/\hbar}\,\hat{g}_{\m_0}\,e^{-i\hat{H}_{\m_0}t/\hbar}
\ket{\Ps^{\rm shad}_{{\vec x}_0}(0)}.
\ee
Therefore, the expectation value $\langle\hat{g}_{\m_0}\rangle_{{\vec x}_0}(t)$ is a constant
if and only if \eqnref{eqn:g and H} holds.

Note that while Consequence~\ref{consequence 1} is simply the result of
\eqnref{eqn:symmetry criterion}, Consequences \ref{consequence 2} and \ref{consequence 3} are different
from it. Either of the consequences could fail while its counterpart
in the standard quantum mechanics holds.

The description of Galileo symmetry is somewhat different from the
above and only the analogue of Consequence~\ref{consequence 1} makes sense because
Galileo boost involves the temporal variable $t$. We will treat it
in a  separate manner in \secref{sec:galileo symmetry}.

\subsection{Translational Symmetry}\label{sec:translational symmetry}
To study the translational symmetry, we study a system with the
potential invariant in some spatial direction specified by the unit
vector ${\vec n}$; i.e. $V({\vec x})=V({\vec x}+\l{\vec n})$ for any
arbitrary $\l$. Classically, we will have the symmetry under the
translation in the ${\vec n}$-direction. Let us check if this system
admits the symmetry in the polymer particle Hilbert space.

In \secref{sec:translation}, we have shown that the translation
operator in ${\vec n}$-direction is $\hat{T}(\m {\vec n})=\opV{-\m
{\vec n}}$ and the action is given by \eqnref{eqn:translation}. It
is easy to show that $\hat{T}(\m {\vec n})$ commutes with
$\widehat{K^2_i}_{\!\m_0}$ and the potential $V(\hat{\vec x})$.
Hence, it admits the translational symmetry in the polymer particle framework
and Consequence~\ref{consequence 1} is true.

Next, let the generator be $\hat{g}_{{\vec n},\m_0}=\hbar \hat{{\vec
K}}_{\m_0}\!\!\cdot{\vec n} \equiv\hbar\sum_{i=1}^3
\hat{{K}}_{i,\m_0}\!\!\cdot n_i$ with $\hat{{K}}_{i,\m_0}$ defined
in \eqnref{eqn:momentum operator}. It can be readily shown that
\eqnref{eqn:g and H} is satisfied and therefore Consequences \ref{consequence 2} and \ref{consequence 3} also hold.

\subsection{Rotational Symmetry}\label{sec:rotational symmetry}
To study the rotational symmetry, consider a system with the axial
symmetry, say, in the $z$-direction; i.e. the potential satisfies
$V({\vec x})=V({\bf R}_z {\vec x})$, where ${\bf R}_z$ is the rotation
about the $z$-axis.

In \secref{sec:rotation}, we have known that the rotation operator acting on
the polymer particle Hilbert space behaves as \eqnref{eqn:rotation}. It is then easy
to show that
\be
\hat{D}({\bf R}_z)^\dagger V(\hat{{\vec x}})\, \hat{D}({\bf R}_z)=V(\hat{{\vec x}})
\ee
and
\be
\hat{D}({\bf R}_z)^\dagger\, \widehat{K^2_z}_{\!\m_0}
\hat{D}({\bf R}_z)=\widehat{K^2_z}_{\!\m_0}.
\ee
On the other hand, denoting
${\bf R}_z\vec{x}=(x\cos\!\ph-y\sin\!\ph,x\sin\!\ph+y\cos\!\ph,z)\equiv(x',y',z)$, we have
\ba\label{eqn:DKD}
&&\hat{D}({\bf R}_z)^\dagger\left(\widehat{K^2_x}_{\!\m_0}+\widehat{K^2_y}_{\!\m_0}\right)
\hat{D}({\bf R}_z)\ket{x,y,z}
=\hat{D}({\bf R}_z)^\dagger\left(\widehat{K^2_x}_{\!\m_0}+\widehat{K^2_y}_{\!\m_0}\right)
\ket{x',y',z}\nn\\
&=&\frac{1}{\m_0^2}\hat{D}({\bf R}_z)^\dagger
\Big\{2\ket{x',y',z}
-\ket{x'+\m_0,y',z}-\ket{x'-\m_0,y',z}\nn\\
&&\qquad +2\ket{x',y',z}-\ket{x',y'
+\m_0,z}-\ket{x',y'-\m_0,z}\Big\}\nn\\
&=&\frac{1}{\m_0^2}
\Big\{4\ket{x,y,z}-\ket{x+\m_0\cos\!\ph,y-\m_0\sin\!\ph,z}
-\ket{x-\m_0\cos\!\ph,y+\m_0\sin\!\ph,z}\nn\\
&&\qquad-\ket{x+\m_0\sin\!\ph,y+\m_0\cos\!\ph,z}
-\ket{x-\m_0\sin\!\ph,y-\m_0\cos\!\ph,z}\Big\},
\ea
which is different from
\ba
&&\left(\widehat{K^2_x}_{\!\m_0}+\widehat{K^2_y}_{\!\m_0}\right)
\ket{x,y,z}\nn\\
&=&\frac{1}{\m_0^2}
\Big\{4\ket{x,y,z}-\ket{x+\m_0,y,z}-\ket{x-\m_0,y,z}
-\ket{x,y+\m_0,z}-\ket{x,y-\m_0,z}\Big\},\qquad
\ea
unless the
rotation angle $\ph$ about $z$-axis is a multiple of $\p/2$, in
which case the rotation transforms the regular lattice $\a^{{\vec
x}_0}$ to itself. Therefore, \eqnref{eqn:symmetry criterion} fails
to be satisfied and Consequence~\ref{consequence 1} does not hold in general!

The problem occurs due to the fact that although the finite rotation
can be carried out in the polymer particle framework, the operator
$\hat{D}({\bf R})$ cannot be decomposed into the fundamental
operators as commented in the end of \secref{sec:rotation}. In
particular, it cannot be generated by $\hat{{\vec x}}$ and
$\opV{\m_0(N_1{\vec e}_1+N_2{\vec e}_2+N_3{\vec e}_3)}$ and spoils
the superselection of each $\hilbert^{{\vec x}_0}_{\rm Poly}$. (See
also the comment in the end of \secref{sec:commutation}.) In a
sense, the continuous rotation $\hat{D}({\bf R})$ is not totally observationally relevant when the operator $\hat{H}_{\m_0}$
modified with $\m_0$ is to govern the ``fundamental dynamics'' on
$\hilbert_{\rm Poly}$.

Let us then investigate Consequences \ref{consequence 2} and \ref{consequence 3}. First, by \eqnref{eqn:L and xi} and
\eqnref{eqn:K square}, we note that
\be
[\hat{L}_{i,\m_0},\widehat{K^2_j}_{\!\m_0}]=0
\ee
and thus $\hat{L}_{i,\m_0}$ commutes with the kinematic part of the Hamiltonian defined
in \eqnref{eqn:hamiltonian}.

On the other hand, by induction based on \eqnref{eqn:L and x}, we have
\be\label{eqn:Lz and x y}
[\hat{L}_{z,\m_0},\hat{x}^n]=i\hbar\,n\,\hat{x}^{n-1}\,\hat{y}\,\hat{\x}_{x,\m_0}
\quad\text{and}\quad
[\hat{L}_{z,\m_0},\hat{y}^n]=-i\hbar\,n\,\hat{y}^{n-1}\,\hat{x}\,\hat{\x}_{y,\m_0}.
\ee
Because of the presence of $\hat{\x}_{i,\m_0}$ in \eqnref{eqn:Lz and x y}, even if $\hat{L}_z$
commutes with the potential in the standard quantum mechanics,
$[\hat{L}_{z,\m_0},V(\hat{x},\hat{y},\hat{z})]$ fails to vanish except for the
case when $V(\hat{x},\hat{y},\hat{z})=V(\hat{z})$.

To estimate how bad the violation of the angular momentum conservation is, we consider a simple
harmonic oscillator with $z$-axial symmetry:
\be\label{eqn:SHO}
\hat{H}_{\m_0}=\frac{\hbar}{2m}\sum_{i=1}^3\widehat{K^2_i}_{\!\m_0}+\frac{1}{2}m\o^2(\hat{x}^2+\hat{y}^2).
\ee
This Hamiltonian gives us
\be
[\hat{L}_{z,\m_0},\hat{H}_{\m_0}]=i\hbar\o^2
\left(\hat{x}\hat{y}\hat{\x}_{x,\m_0}-\hat{y}\hat{x}\hat{\x}_{y,\m_0}\right).
\ee
By \eqnref{eqn:time change of g}, the time variation of the expectation value of $\hat{L}_{z,\m_0}$
is given by
\ba
&&\frac{d}{dt}\langle\hat{L}_{z,\m_0}\rangle_{{\vec x}_0}(t)=
-\frac{i}{\hbar}\round{\Ps(t)}
[\hat{L}_{i,\m_0},\hat{H}_{\m_0}]
\ket{\Ps^{\rm shad}_{{\vec x}_0}(t)}\nn\\
&=&m\o^2
\round{\Ps(t)}
\left(\hat{x}\hat{y}\hat{\x}_{x,\m_0}-\hat{y}\hat{x}\hat{\x}_{y,\m_0}\right)
\ket{\Ps^{\rm shad}_{{\vec x}_0}(t)}\nn\\
&=&m\o^2
\round{\Ps(t)}
\left(\hat{x}\hat{y}\left(1-\m_0^2\widehat{K^2_x}_{\!\m_0}/2\right)
-\hat{y}\hat{x}\left(1-\m_0^2\widehat{K^2_y}_{\!\m_0}/2\right)\right)
\ket{\Ps^{\rm shad}_{{\vec x}_0}(t)}\nn\\
&\sim& m\o^2{\cal O}(k^2\m_0^2)
\round{\Ps(t)}\hat{x}\hat{y}\ket{\Ps^{\rm shad}_{{\vec x}_0}(t)}
\sim m\o^2\langle\hat{x}\hat{y}\,\rangle{\cal O}(k^2\m_0^2),
\ea
where \eqnref{eqn:K square} has been used and $\langle\hat{x}\hat{y}\rangle:=\bra{\Ps(t)}\hat{x}\hat{y}\ket{\Ps(t)}$ is the expectation value of $\hat{x}\hat{y}$ in the standard quantum mechanics.
The result is non-vanishing but highly suppressed if $k\m_0\ll1$. For
$k\m_0\gtrsim 1$, the violation of the angular momentum conservation
is noticeable, but again the relativistic correction needs to be
taken into account long before such high momentum is reached.

By introducing the discreteness $\m_0$ to the momentum operator in
\eqnref{eqn:momentum operator} and to the Hamiltonian in
\eqnref{eqn:hamiltonian} and \eqnref{eqn:K square}, we explicitly
break the rotational symmetry. Since the modified Hamiltonian is
used for the ``fundamental dynamics'' on $\hilbert_{\rm Poly}$,
Consequence~\ref{consequence 1} can no longer hold exactly even for a free particle.
Nevertheless, we can still probe the associated angular momentum and
the conservation law still makes sense with high accuracy in the low
energy regime.

\subsection{Galileo Symmetry}\label{sec:galileo symmetry}
When we say a dynamical system respects Galileo symmetry, it means
two reference frames $F$ and $F'$ with a constant relative velocity
are physically equivalent. In the standard quantum mechanics, that
means a state evolved by $t$ and then followed by a boost
instantaneously is the same as the state boosted at time zero and
followed by the evolution with $t$. More precisely, that is
\be\label{eqn:galileo symmetry}
\hat{G}({\vec v},t)\hat{U}_F(t,0)\ket{\ps}=\hat{U}_{F'}(t,0)\hat{G}({\vec v},0)\ket{\ps}
\qquad\text{for all $\ket{\ps}$},
\ee
where $\hat{U}_F(t',t)$ is the time evolution operator based on the Hamiltonian $\hat{H}_F(t)$
viewed in the reference $F$; the Hamiltonians
are\footnote{Here, we adopt the convention
that the transformation operator acts on the states instead of the observable operators; i.e.
$\ket{\Ps}\rightarrow\ket{\Ps'}=\hat{G}({\vec v},t)\ket{\Ps}$ while
$\hat{{\vec x}}$ and $\hat{{\vec p}}$ are kept the same for the two frames. However, The potential
experienced by the moving frame $F'$ is different from that by $F$.}
\be\label{eqn:HF}
\hat{H}_F(t)=\frac{\widehat{{\vec p}^{\,2}}}{2m}+V(\hat{{\vec x}})
\qquad\text{for frame }F
\ee
and
\be\label{eqn:HF'}
\hat{H}_{F'}(t)=\frac{\widehat{{\vec p}^{\,2}}}{2m}+V(\hat{{\vec x}}-{\vec v}t)
\qquad\text{for frame }F'.
\ee

Taking the time derivative on both sides of \eqnref{eqn:galileo symmetry} yields
\be\label{eqn:galileo symmetry2}
\left(\frac{d}{dt}\hat{G}({\vec v},t)\right)\hat{U}_F(t,0)
-\frac{i}{\hbar}\hat{G}({\vec v},t)\hat{H}_F(t)\hat{U}_F(t,0)
=-\frac{i}{\hbar}\hat{H}_{F'}(t)\hat{U}_{F'}(t,0)\hat{G}({\vec v},0).
\ee
By \eqnref{eqn:galileo transform} (or \eqnref{eqn:G} for both the standard and polymer particle
frameworks), we have
\be
\frac{d}{dt}\hat{G}({\vec v},t)=
\left\{\frac{i\,mv^2}{2\hbar}+\left(\frac{d}{dt}\opV{-{\vec v}t}\right)\hat{V}^\dagger(-{\vec v}t)
\right\}\hat{G}({\vec v},t),
\ee
and then imposing
$\hat{U}_{F'}(t,0)\hat{G}({\vec v},0)\hat{U}_{F}^\dagger(t,0)\hat{G}^\dagger({\vec v},t)=1$
(from \eqnref{eqn:galileo symmetry}) on \eqnref{eqn:galileo symmetry2}, we can show
\be\label{eqn:galileo symmetry3}
-\frac{1}{2}mv^2+i\hbar\left(\frac{d}{dt}\opV{-{\vec v}t}\right)\hat{V}^\dagger(-{\vec v}t)
+\hat{G}({\vec v},t)\hat{H}_F(t)\,\hat{G}^\dagger({\vec v},t)
=\hat{H}_{F'}(t).
\ee
(In fact, conversely, \eqnref{eqn:galileo symmetry3} also
implies \eqnref{eqn:galileo symmetry}. This can be
proven by the fact that $\hat{U}(t,0)=\mathcal{T}e^{-i\int_0^t dt\hat{H}(t)/\hbar}$ and
the uniqueness theorem of the differential equation.)

In the standard quantum mechanics, the momentum operator is well-defined and
we have $\opV{{\vec v}t}=e^{i\hat{{\vec p}}\cdot{\vec v}t/\hbar}$,
$\hat{G}({\vec v},t)\,\hat{{\vec p}}\,\hat{G}^\dagger({\vec v},t)= \hat{{\vec p}}-m\vec{v}$
and
$\hat{G}({\vec v},t)\,\hat{{\vec x}}\,\hat{G}^\dagger({\vec v},t)= \hat{{\vec x}}-\vec{v}t$.
\eqnref{eqn:galileo symmetry3} then implies
\be\label{eqn:galileo symmetry4}
\hat{H}_{F'}(t)=\left.\hat{H}_F(t)
\right|_{\hat{{\vec x}}\rightarrow\hat{{\vec x}}-\vec{v}t,\,\hat{{\vec p}}\rightarrow \hat{{\vec p}}-m\vec{v}}
+\hat{{\vec p}}\cdot {\vec v}-\frac{1}{2}mv^2.
\ee
It is obvious to see that \eqnref{eqn:HF} and \eqnref{eqn:HF'}
satisfy the condition of \eqnref{eqn:galileo symmetry4}. That means
most systems respect Galileo symmetry as long as their Hamiltonians
are of the standard form (kinematic energy plus an arbitrary
potential).

To study the Galileo symmetry in the polymer particle framework, due to absence of the momentum operator,
we have to first replace the Hamiltonians in \eqnref{eqn:HF} and \eqnref{eqn:HF'} with the modified form
of \eqnref{eqn:hamiltonian}; that is
\be\label{eqn:HFmu}
\hat{H}_{F,\m_0}(t)=\frac{\hbar^2}{2m}
\left(\widehat{K^2_x}_{\!\m_0}+\widehat{K^2_y}_{\!\m_0}+\widehat{K^2_z}_{\!\m_0}\right)+V(\hat{{\vec x}})
\qquad\text{for frame }F
\ee
and
\be\label{eqn:HF'mu}
\hat{H}_{F'\!,\m_0}(t)=\frac{\hbar^2}{2m}
\left(\widehat{K^2_x}_{\!\m_0}+\widehat{K^2_y}_{\!\m_0}+\widehat{K^2_z}_{\!\m_0}\right)+V(\hat{{\vec x}}-{\vec v}t)
\qquad\text{for frame }F'.
\ee
Second, in the same spirit, we also adopt the prescription for the time derivative of $\hat{V}$:
\be\label{eqn:dVdt}
\left(\frac{d}{dt}\opV{-{\vec v}t}\right)\hat{V}^\dagger(-{\vec v}t)
\longrightarrow-i\hat{\vec K}_{\m_0}\!\!\cdot{\vec v}
\equiv
-i\left(\hat{K}_{x,\m_0}v_x+\hat{K}_{y,\m_0}v_y+\hat{K}_{z,\m_0}v_z\right).
\ee
But this is equivalent to modify the Galileo transformation operator $\hat{G}({\vec v},t)$
in \eqnref{eqn:G} to
\be\label{eqn:modified galileo transformation}
\hat{G}({\vec v},t)\longrightarrow\hat{G}_{\m_0}({\vec v},t):=
e^{\frac{i}{\hbar}\frac{mv^2}{2}t}\ \hat{V}_{\m_0}(-\vec{v}t)\,\opU{m\vec{v}/\hbar}
\ee
with the operator $\hat{V}_{\m_0}$ regularized from $\hat{V}$:
\be\label{eqn:regularized V}
\opV{-{\vec v}t}\longrightarrow\hat{V}_{\m_0}(-{\vec v}t):=
e^{-i\hat{{\vec K}}_{\m_0}\!\cdot\widetilde{{\vec v}t}}\ \opV{-[{\vec v}t]},
\ee
where $[v_i t]:=\max\left\{n\m_0|n\in{\mathbb Z},n\m_0\leq v_i
t\right\}$ and the remainder $\widetilde{v_i t}:=v_it-[v_it]$.
Because we require \eqnref{eqn:galileo symmetry} to be true for any
$t$, we have to specify $\hat{G}({\vec v},t+\D t)$ for the
infinitesimal $\D t$; while the infinitesimal transformation of
$\hat{V}$ is ill-defined, we regularize it by
\eqnref{eqn:regularized V} and this prescription correspondingly
gives rise to \eqnref{eqn:dVdt}.\footnote{To give \eqnref{eqn:dVdt},
we can simply regularize $\hat{V}(-{\vec v}t)$ to
$\hat{V}_{\m_0}(-{\vec v}t):= e^{-i\hat{{\vec
K}}_{\m_0}\!\cdot{{\vec v}t}}$ rather than \eqnref{eqn:regularized
V}. However, this naive prescription leads to $\hat{G}_{\m_0}({\vec
v},t)\, \hat{{\vec x}}\, \hat{G}_{\m_0}^\dagger({\vec
v},t)=\hat{{\vec x}}-\hat{{\vec \x}}_{\m_0}{{\vec v}t}$, which
causes growing discrepancy with $t$ when compared to the result of
the standard quantum mechanics. With the tamed regularization
instead, \eqnref{eqn:regularized V} gives \eqnref{eqn:G and x} and
the deviation in $x_i$ is
$(1-\hat{\x}_{i,\m_0})\widetilde{v_it}\lesssim\m_0{\cal
O}(k^2\m_0^2)$, which is highly negligible in the low-energy regime.}

Notice that we have
\be\label{eqn:G and V}
\hat{G}_{\m_0}({\vec v},t)\opV{\m_0{\vec e}_i}\,\hat{G}_{\m_0}^\dagger({\vec v},t)=
\hat{G}({\vec v},t)\opV{\m_0{\vec e}_i}\,\hat{G}^\dagger({\vec v},t)
=e^{-\frac{i}{\hbar}m\m_0v_i}\opV{\m_0{\vec e}_i}
\ee
and
\be\label{eqn:G and x}
\hat{G}_{\m_0}({\vec v},t)\, \hat{{x_i}}\,\hat{G}_{\m_0}^\dagger({\vec v},t)
=\hat{{x_i}}-[{v_i}t]-\hat{{\x}}_{i,\m_0}\widetilde{{v_i}t}.
\ee
To prove \eqnref{eqn:G and x}, we have used $\opV{-[v_i t]}\,\hat{x}_j\hat{V}^\dagger(-[v_i t])=\hat{x}_j-\d_{ij}[v_it]$
and $e^{-i\hat{K}_{i\m_0}\widetilde{v_it}}\,\hat{x}_j\, e^{i\hat{K}_{i\m_0}\widetilde{v_it}}
=\hat{x}_j-\d_{ij}\widetilde{v_it}$
(by \eqnref{eqn:x and K} and the identity $[\hat{B},e^{x\hat{A}}]=e^{x\hat{A}}[\hat{B},\hat{A}]x$).
(Also note that \eqnref{eqn:G and V} means Galileo transformation
and translation commute geometrically,
but there is an irremovable phase factor quantum mechanically.)

The condition for the Galileo symmetry \eqnref{eqn:galileo symmetry3} is now modified to
\be\label{eqn:galileo symmetry5}
-\frac{1}{2}mv^2+\hbar\,\hat{\vec K}_{\m_0}\!\!\cdot{\vec v}
+\hat{G}_{\m_0}({\vec v},t)\,\hat{H}_{F,\m_0}(t)\,\hat{G}_{\m_0}^\dagger({\vec v},t)
=\hat{H}_{F'\!,\m_0}(t).
\ee
With \eqnref{eqn:G and V} and \eqnref{eqn:G and x}, the left-hand side of \eqnref{eqn:galileo symmetry5}
becomes
\ba\label{eqn:left hand side}
&&-\frac{1}{2}mv^2+\hbar\,\hat{\vec K}_{\m_0}\!\!\cdot{\vec v}
+\frac{\hbar^2}{2m\m_0^2}\sum_{i=1}^{3}
\left[2-e^{-\frac{i}{\hbar}m\m_0v_i}\opV{\m_0\vec{e}_i}
-e^{\frac{i}{\hbar}m\m_0v_i}\opV{-\m_0\vec{e}_i}\right]\nn\\
&&\quad+V(\hat{{\vec x}}-[{\vec v}t]-\hat{{\vec \x}}_{\m_0}\widetilde{{\vec v}t}),
\ea
which can be expressed as
\ba\label{eqn:left hand side2}
&&
-\frac{1}{2}\sum_{i=1}^{3}mv_i^2\left(1-\hat{\x}_{i,\m_0}\right)
+\frac{\hbar^2}{2m\m_0^2}\sum_{i=1}^{3}
\left[2-\opV{\m_0\vec{e}_i}
+
-\opV{-\m_0\vec{e}_i}\right]\nn\\
&&\qquad
+\frac{\hbar^2}{m\m_0^2}\ {\cal O}\left(\left(\frac{m\m_0 v}{\hbar}\right)^3\right)
+V(\hat{{\vec x}}-{\vec v}t)
+\widetilde{{\vec v}t}\cdot\!\vec{\nabla}V\,{\cal O}(k^2{\m_0^2})\nn\\
&=&\frac{\hbar^2}{2m\m_0^2}\sum_{i=1}^{3}
\left[2-\opV{\m_0\vec{e}_i}
-\opV{-\m_0\vec{e}_i}\right]
+V(\hat{{\vec x}}-{\vec v}t)\nn\\
&&\quad+\frac{\hbar^2}{m\m_0^2}\ {\cal O}\left(\left(\frac{m\m_0 v}{\hbar}\right)^3\right)
+\left(\frac{1}{2}mv^2+\m_0|\vec{\nabla} V|\right)\,{\cal O}(k^2{\m_0^2}),
\ea
where $\vec{\nabla} V$ is the gradient of potential and
$\hbar\vec{k}$ is the momentum of the moving particle in the frame
$F$. The result of \eqnref{eqn:left hand side2} agrees with the
right-hand side of \eqnref{eqn:galileo symmetry5} (i.e.
\eqnref{eqn:HF'mu}) with minuscule discrepancy. In conclusion, the
Galileo symmetry in the polymer particle framework is still respected by most
systems, but only approximately; however, the disagreement is highly
negligible as long as $m\m_0v/\hbar\ll 1$ and $k\m_0\ll 1$. When the
boost velocity $v$ or the momentum of the particle $k$ is too big,
the Galileo symmetry does not hold anymore, but again the
relativistic effect should take place far before the
breakdown.\footnote{We can adopt the different perspective to say
that when we modify the Hamiltonian in the frame $F$ from
\eqnref{eqn:HF} to \eqnref{eqn:HFmu}, correspondingly, we should
modified the Hamiltonian in the moving frame $F'$ as
\eqnref{eqn:left hand side} instead of \eqnref{eqn:HF'mu}. If we
take \eqnref{eqn:left hand side} as the viable Hamiltonian for $F'$,
the Galileo symmetry will be respected exactly. However, this
prescription is difficult to be justified.}

The pathological trait in the above investigation is the \textit{ad
hoc} modification in \eqnref{eqn:modified galileo transformation}
and \eqnref{eqn:regularized V}. This procedure is necessary because
we need the time derivative of $\opV{-{\vec v}t}$ in
\eqnref{eqn:galileo symmetry3}. While the fundamental dynamics is
governed by the new Hamiltonians \eqnref{eqn:HFmu} and
\eqnref{eqn:HF'mu}, the time $t$ is not discretized. However, this
problem can be circumvented if we interchange the roles of
$\hat{{\vec x}}$ and $\hat{{\vec p}}$ and thus impose the
``fundamental momentum'' $\l_0$ (which is of dimension
$[\text{length}]^{-1}$) in $k$-space instead of $\m_0$ in $x$-space.
The momentum operator is well-defined and $\opV{{\vec \m}}=e^{i{\vec
\m}\cdot\hat{{\vec p}}/\hbar}$.\footnote{The opposite ways of treating the canonical pair of variables in the phase space are sometimes referred as different choices of ``polarization'' by some authors.}

Following the same procedures as above, we introduce the modified position operator
\be
\hat{X}_{i,\l_0}:=\frac{1}{2\l_0i}\left(\opU{\l_0{\vec e}_i}-\opU{-\l_0{\vec e}_i}\right).
\ee
Accordingly, the Hamiltonians in the frames $F$ and $F'$ are given by
\be
\hat{H}_{F,\l_0}(t)=\frac{\widehat{{\vec p}^{\,2}}}{2m}+V(\hat{{X}}_{i,\l_0})
\ee
and
\be\label{eqn:HF'lam}
\hat{H}_{F',\l_0}(t)=\frac{\widehat{{\vec p}^{\,2}}}{2m}+V(\hat{{X}}_{i,\l_0}-{v_i}t).
\ee
Since time derivative of $\hat{V}$ is well-defined:
$\frac{d}{dt}\opV{-{\vec v}t}=-i\frac{{\vec v}\cdot{\vec p}}{\hbar}\,\opV{-{\vec v}t}$, we do not need
to modify the operator $\hat{G}({\vec v},t)$ and it keeps the same as \eqnref{eqn:G}.
A simple calculation shows
$\hat{G}({\vec v},t)\,\hat{{\vec p}}\,\hat{G}^\dagger({\vec v},t)=\hat{{\vec p}}-m{\vec v}$
and
$\hat{G}({\vec v},t)\,\opU{\l_0{\vec e}_i}\,\hat{G}^\dagger({\vec v},t)
=e^{-i\l_0v_it}\,\opU{\l_0{\vec e}_i}$, which leads to
\be
\hat{G}({\vec v},t)\,\hat{X}_{i,\l_0}\,\hat{G}^\dagger({\vec v},t)=
\frac{e^{-i\l_0v_it}\opU{\l_0{\vec e}_i}-e^{i\l_0v_it}\opU{-\l_0{\vec e}_i}}{2\l_0i}.
\ee

Put all together, the left-hand side of \eqnref{eqn:galileo symmetry3} becomes
\ba
&&-\frac{1}{2}mv^2+\vec{p}\cdot\vec{v}+\frac{(\vec{p}-m\vec{v})^2}{2m}
+V\left(\frac{e^{-i\l_0v_it}\opU{\l_0{\vec e}_i}-e^{i\l_0v_it}\opU{-\l_0{\vec e}_i}}{2\l_0i}\right)\nn\\
&=&
\frac{\vec{p}^{\,2}}{2m}+V\left(\hat{X}_{i,\l_0}-v_it\right)
+\vec{v}t\cdot\!\vec{\nabla}V\,{\cal O}(\l_0^2x^2),
\ea
which in general does not agree with the right-hand side of \eqnref{eqn:galileo symmetry3}
(i.e \eqnref{eqn:HF'lam}) if the spatial position of interest ($\vec{x}$ or $\vec{x}+\vec{v}t$)
is distant from the origin.
In this alternative formulation, the Galileo symmetry holds only if the space range is confined in $x\ll\l_0^{-1}$.

In fact, exchanging the roles of $\vec{x}$ and $\vec{p}$ causes a more general problem.
In the alternative framework, the transformations in the polymer particle representation fails to
consistently reproduce their counterparts in the standard quantum mechanics. To see this,
we perform the exchanges
$x\leftrightarrow k$, $\m_0\leftrightarrow\l_0$ and so on on
\eqnref{eqn:x expectation} and \eqnref{eqn:p expectation}, which then give
\be\label{eqn:p expectation2}
\langle\hat{p}_i\rangle_{\vec \z}:=
\frac{\round{\Ps_{\vec \z}}\hat{p}_i\ket{\Ps_{{\vec \z},\g}^{\rm shad}}}{\norm{\Ps_{{\vec \z},\g}^{\rm shad}}^2}
=p_i\left[1+{\cal O}\left(e^{-\frac{\p^2\ell^2}{d^2}}\right)\right]
\ee
and
\be\label{eqn:x expectation2}
\langle\hat{x}_i\rangle_{\vec \z}=
\frac{\round{\Ps_{\vec \z}}\hat{X}_{i,\l_0}\ket{\Ps_{{\vec \z},\g}^{\rm shad}}}{\norm{\Ps_{{\vec \z},\g}^{\rm shad}}^2}
=x_i\left[1+{\cal O}\left(x^2\l_0^2\right)+{\cal O}\left(\frac{d^2}{\ell^2}\right)\right],
\ee
where $1/d$ is the spread width in $k$-space of the coherent state and $1/\ell$ is the spacing of the regular
lattice $\g$ which is embedded in $k$-space and $1/\ell\ll1/d$.
We notice that \eqnref{eqn:x expectation2} can be far deviated from $x_i$ if $x\gtrsim\l_0$
and this discrepancy cannot attribute to relativistic effect.

The above discussion suggests that $\vec{x}$ and $\vec{p}$ are not
on equal footing: To construct the polymer particle description, we should
keep $\hat{\vec{x}}$ while dismiss $\hat{\vec{p}}$ and replace it by
$\hat{V}(\vec{\m})$, not the other way around. It seems to be the
non-relativistic physics that discriminates $\vec{x}$ and $\vec{p}$
and prevents the interchange. In LQG, perhaps the non-relativistic
limit also plays a role in differentiating the canonical pair of
$SU(2)$ connections ${A^i}_a$ and densitized triads ${E_i}^a$.

Furthermore, since interchanging the roles of $\vec{x}$ and
$\vec{p}$ does not work, we are forced to modify the Galileo
transformation operator as in \eqnref{eqn:modified galileo
transformation} with the regularization in \eqnref{eqn:regularized
V}. This result tells that, even though we treat $t$ as a
continuous variable, the moving frame $F'$ experiences the discreteness in time due to the irremovable regularization in accordance with the fundamental length $\m_0$. In a sense, the spatial discreteness gives rise
to the temporal discreteness when probed by a boosted reference. At
this moment, it is unclear whether this is only an artifact in the
simple toy model or this observation can be generalized to the
full theory of quantum geometry.

\section{Conclusion}\label{sec:conclusion}
The complete Galileo transformations and symmetries are investigated
in the polymer particle representation of quantum mechanics of a
non-relativistic particle constructed in \cite{Ashtekar:2002sn}. By
exploiting the fact that the standard coherent state in $\sw$
uniquely gives a polymer coherent state in $\dual{\cyl}$, all the
finite Galileo transformations (translation, rotation and Galileo
boost) can be naturally defined in the polymer particle Hilbert
space. Furthermore, the spatial translation and Galileo boost can be
expressed in terms of the fundamental operators $\opU{\vec{\l}}$ and
$\opV{\vec{\m}}$, but the rotational operator cannot.

Three different consequences of continuous symmetries are explored
in detail. It shows that in the polymer particle framework, these
three consequences do not necessarily imply to one another. If a
system is translationally symmetric in the standard quantum
mechanics, all three consequences of the translational symmetry also
hold in the polymer particle Hilbert space with the modified
Hamiltonian $\hat{H}_{\m_0}$. For the case of rotational symmetry,
Consequences \ref{consequence 1} and \ref{consequence 2} no longer hold in the polymer particle
description due to the fact that the rotational symmetry is
explicitly broken when the fundamental discreteness is imposed.
Nevertheless, the change rate of the angular momentum is highly
suppressed in the low energy regime and thus Consequence~\ref{consequence 3} is still
respected with negligible deviation. Finally, Galileo symmetry is
treated separately and shown to be an excellent approximation in the
polymer particle framework as long as the momentum and boost
velocity are small enough.

As with the Galileo symmetry, it also suggests that non-relativistic
physics may explain why $\vec{x}$ and $\vec{p}$ are not on equal
footing in the polymer particle representation. Furthermore, we also
observe that the discreteness in space may give rise to the temporal
discreteness when probed by a boosted reference. Further
investigations in more sophisticated models are necessary for this
observation.

\begin{acknowledgments}
The author is grateful for the warm hospitality during his visit at
IGPG and thanks Abhay Ashtekar for initiating this work and useful
discussions.
\end{acknowledgments}


\end{document}